# Tailoring the spatio-temporal distribution of diffractive focused ultrashort pulses through pulse shaping

BENJAMÍN ALONSO,[1,2,*] JORGE PÉREZ-VIZCAÍNO,[3] GLADYS MÍNGUEZ-VEGA,[3] AND ÍÑIGO J. SOLA[1]

[1] *Grupo de Aplicaciones del Láser y Fotónica (ALF), Departamento de Física Aplicada, University of Salamanca, E37008 Salamanca, Spain*
[2] *Sphere Ultrafast Photonics, S.A., Parque de Ciência e Tecnologia da Universidade do Porto, R. do Campo Alegre 1021, Edifício FC6, 4169-007 Porto, Portugal*
[3] *GROC UJI, Institute of New Imaging Technologies, Universitat Jaume I, 12071 Castellón, Spain*
*\* b.alonso@usal.es*

**Abstract:** Focusing control of ultrashort pulsed beams is an important research topic, due to its impact to subsequent interaction with matter. In this work, we study the propagation near the focus of ultrashort laser pulses of ~25 fs duration under diffractive focusing. We perform the spatio-spectral and spatio-temporal measurements of their amplitude and phase, complemented by the corresponding simulations. With them, we demonstrate that pulse shaping allows modifying in a controlled way not only the spatio-temporal distribution of the light irradiance in the focal region, but also the way it propagates as well as the frequency distribution within the pulse (temporal chirp). To gain a further intuitive insight, the role of diverse added spectral phase components is analyzed, showing the symmetries that arise for each case. In particular, we compare the effects, similarities and differences of the second and third order dispersion cases.

## 1. Introduction

Ultrafast and intense Optics has become a major tool in different aspects of Science (Biophotonics, Physics, Chemistry, etc.). Because of the short duration of the pulses, light can be used as a probe to measure features of extremely fast dynamics in Nature (atomic and molecular dynamics, establishment of chemical bonds, etc.), or also to obtain high power and high intensity sources to trigger highly nonlinear processes (e.g., material ablation, high-order harmonic generation (HHG), laser driven particle acceleration…). To obtain shorter and more intense pulses at their focus, in general an aberration-free focusing approach must be employed to avoid a larger focal spot size and/or longer pulse durations, as well as undesired spatio-temporal coupling.

However, there is also an increasing number of examples where the full potential of the laser source can be optimized by changing the spatial or temporal (or both) characteristics of the beam with respect to the *ideal* focusing (i.e., diffraction limited spot in the spatial domain, Fourier-transform limited pulses in the temporal domain and absence of spatio-temporal coupling). In this way, it has been reported that some *non-ideal* focusing schemes improve the results in applications. For example, in the case of flat-top focal distributions in HHG [1], spatio-temporal couplings (e.g., with transversal spatial chirp to create attosecond lighthouses [2], angular chirp for angle-dependent emission of HHG [3], space-time focusing for microprocessing [4], or to control light velocity [5]). Also, beam aberrations are used to enhance nonlinear confocal microscopy [6], the filamentation process [7], or to generate a uniform ellipsoidal particle distribution [8]. Temporal pulse shaping [9] has applications in microprocessing, Biomedicine, nonlinear Optics, etc. Finally, the control of the carrier-envelope phase (CEP) when the pulses are focused [10], including the effect of the chirp [11], also influences CEP-sensitive processes.

In this context, diffractive optical elements (DOEs) have been shown to be very flexible components to modify the spatial or temporal characteristics of the laser beam. For example, they have been used in achromatic multibeam elements [12], chromatic focusing to tune second harmonic generation (SHG) [13] or to perform chromatic SHG imaging [14], and to create light fractals [15]. Also, phase masks may be used to enhance filament characteristics [16], or in diffractive pulse shapers [17], among others.

In contrast to the behavior of monochromatic light, ultrashort pulses (thus with broadband spectra) exhibit complex spatio-temporal structures when propagating after a DOE, since wavelength dependent diffraction originates certain spatio-temporal patterns, as shown for example in [18,19]. Indeed, since light is pulsed, this pattern depends on the phase of its different wavelength components, i.e., the spectral phase of the pulse. Therefore, the control of the input pulse spectral phase (e.g., by means of pulse shaping techniques [20]) turns into an additional control parameter in order to modify the spatio-temporal pattern in the focal region. As an example of its applications, in previous works [21,22], Li and co-workers have shown that, through superposition of chromatic aberration from refractive focusing (singlet lens) and input pulse shaping by an acousto-optic programmable dispersive filter (AOPDF) [23], it is possible to tailor time evolving beams at their focal plane. In particular, they use uniform ellipsoidal beams to initiate X-ray free electron lasers (XFEL) beams [8].

In this work, we explore the impact of input pulse shaping on the distribution of light after being focused by a DOE (a kinoform diffractive lens, DL [18,24]). Our aim is to analyze the beam behavior at different planes around the focal region when input spectral phase is changed, and to understand the role of its main components (Group Dispersion Delay, GDD; and Third Order Dispersion, TOD) on the final spatio-temporal distribution. This will lead to the observation of a number of symmetries governing the spatial and temporal interplay along the focusing region, as well as to the finding of different distributions of the spatio-temporal irradiance. Firstly, we present the theoretical and experimental methods, with the description of the elements and parameters used in this work. Then, we show the behavior of the beam propagation in the spatio-spectral domain. After that, we study the experimental and simulated spatio-temporal structures of the beam at different propagation distances along the focal region and the corresponding instantaneous frequencies (chirp). In order to gain a further insight, in Section 4 (Discussion) we analyze the role of some different added spectral phase components as a control parameter (namely the GDD and that TOD) for a better understanding of their effect in the final outcome, as well as to theoretically demonstrate the symmetries observed in the experimental and simulated results.

## 2. Theoretical and experimental methods

Theoretical simulations were implemented modeling the DL to present a focal length $f$ inversely proportional to the wavelength $\lambda$, given by $f(\lambda) = \lambda_0 f_0 / \lambda$, being in our case $\lambda_0 = 785\,nm$, the central wavelength of the laser, and $f_0 = 108.0\,mm$, the focal for $\lambda_0$. As shown in [18], this simplified model is enough to accurately simulate the light behavior when propagating after the DL. In the spatial domain, we used a Gaussian profile with 7-mm diameter (full width at $1/e^2$), and we simulated the input Gaussian spectrum with FWHM (full-width at half-maximum) of 35 nm, both values directly taken from the experimental conditions. Thanks to the radial symmetry of the DL, we calculated Fresnel diffraction expressed in cylindrical coordinates [25] for each wavelength component of the beam. This allows working out the spatio-temporal distribution of the pulse by inverse Fourier-transforming the diffracted pulse. The propagation dynamics is obtained by calculating the diffraction pattern for different propagation distances around the focal region.

The light source used in the experiments was a Ti:Sapphire chirped pulse amplification laser system (Femtolasers FemtoPower Compact Pro HE CEP), delivering pulses centered at

785 nm, with a bandwidth of 35 nm (FWHM) and a duration of 24 fs (FWHM) for Fourier-limit conditions, working at a repetition rate of 1 kHz. The pulse shaping was performed introducing a given dispersion by means of a programmable acousto-optic device (Dazzler [26], from Fastlite). Then, the beam was chromatically focused with the DL. The spatio-temporal characterization was performed using the STARFISH technique [27], which uses a fiber optic coupler to measure the spatially-resolved spectral interferometry of the unknown pulse with a known reference pulse. The beam was retrieved both, experimentally and theoretically, in a range of $\delta z = \pm 5\,mm$ around the DL focal plane located at a propagation distance $z_0 = f_0 = 108.0\,mm$, corresponding to the central wavelength $\lambda_0 = 785\,nm$. Therefore, not only the focal plane is characterized, but also the surrounding positions, providing the evolution of the spatio-temporal pattern of the pulse near the focus. To scan this propagation, the distance between the DL and the collection fiber arm was varied by translating the DL along the optical axis. The reference pulse required for the retrieval of the spectral phase from the interferences was characterized using an auxiliary temporal/spectral measurement (a FROG device [28]).

Concerning the study of the effect of the input spectral phase over the spatio-temporal irradiance pattern at focus (or near focus), we have considered the spatio-temporal distributions arising when scanning separately the GDD from $-3000\,fs^2$ to $+3000\,fs^2$, and the TOD from $-50000\,fs^3$ to $+50000\,fs^3$, both in the simulations and the experiments.

## 3. Results

The focal length dependence $f(\lambda)$ causes that longer wavelengths focus before than shorter ones. This effect modulates the beam spectrum along propagation around the central focal plane [18], as shown in Fig. 1 (the same logarithmic scale bar is used for Fig. 1–5), which is directly related to the spatio-temporal patterns presented below. Also, the spectral distributions before and after $z = f_0$ are inverted with respect to the central wavelength if the input spectrum is also symmetrical with respect to $\lambda_0$.

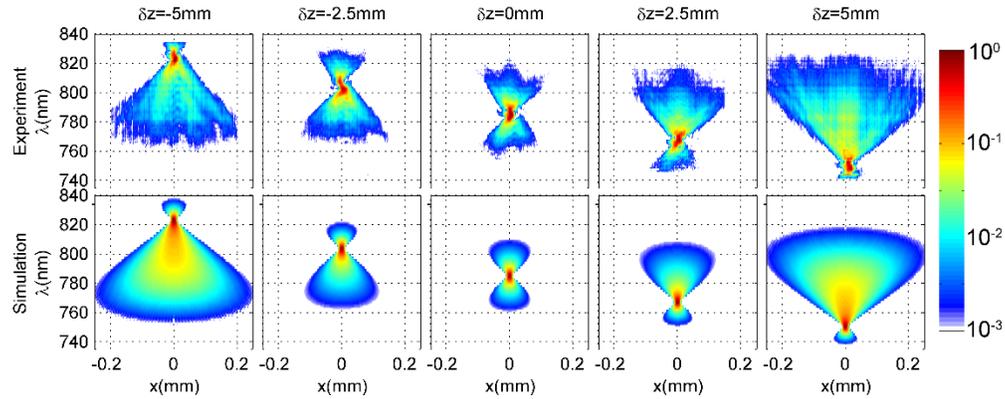

Fig. 1. Experimental (above) and simulated (below) spatio-spectral distribution depending on the propagation plane z, where δz=0 mm corresponds to the focal plane $z = f_0$, the focus for the central wavelength ($\lambda_0$=785 nm). Results are normalized and shown in logarithmic scale, where the scale bar indicates the 3 orders of magnitude represented.

As shown in the demonstration of Eq. (4) (Section 4), this result is exact for the simulated Gaussian spectrum, while it is approximated for the slightly non-symmetrical experimental spectrum. This is also related to the spatio-temporal symmetries found when adding input GDD and TOD to the pulse.

Then, the input pulse GDD was systematically changed and the resulting spatio-temporal irradiance distribution was measured for several planes along the DL focal region. Fig. 2 shows a selection of the results: the columns stand for each value of the input pulse GDD, whereas rows correspond to each plane of propagation at z=f$_0$+δz (before, at and after focus).

The input pulse $GDD = 0\,fs^2$ (central column) corresponds to the conditions studied in [18], (i.e., no spectral phase is added before the DL). When diverse values of input pulse GDD are added, the spatio-temporal irradiance patterns are modified. In particular, an interesting feature is the relation between cases at different $\delta z$ and $GDD$ conditions. Indeed, when analyzing Fig. 2, the different distributions exhibit a symmetry with respect to the δz=0 mm and $GDD = 0\,fs^2$ case (for example, when comparing $\delta z = -5\,mm$ and $GDD = -3000\,fs^2$ with $\delta z = +5\,mm$ and $GDD = +3000\,fs^2$, or $\delta z = +5\,mm$ and $GDD = -3000\,fs^2$ with $\delta z = -5\,mm$ and $GDD = +3000\,fs^2$).

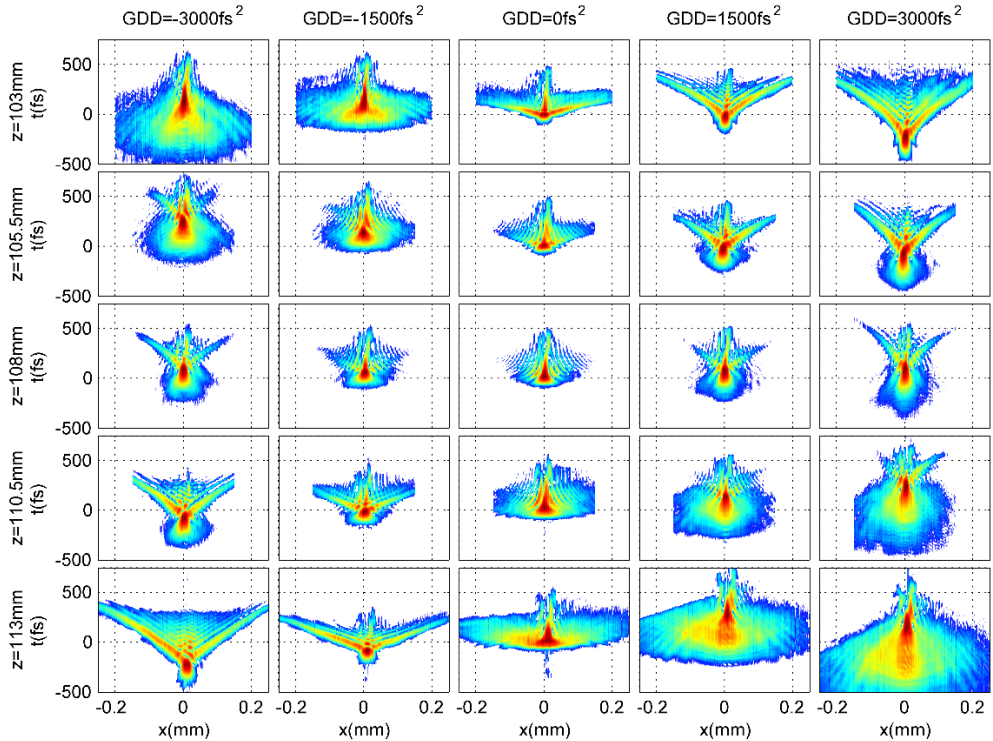

Fig. 2. Experimental spatio-temporal irradiance distribution depending on the propagation plane z (rows) and TOD of the input pulse (columns). δz=0 mm corresponds to $\lambda_0$ focal plane. Results are normalized and shown in logarithmic scale, representing 3 orders of magnitude.

This symmetry is confirmed with the corresponding simulations (Fig. 3), showing that the agreement between theory and experiments is very good.

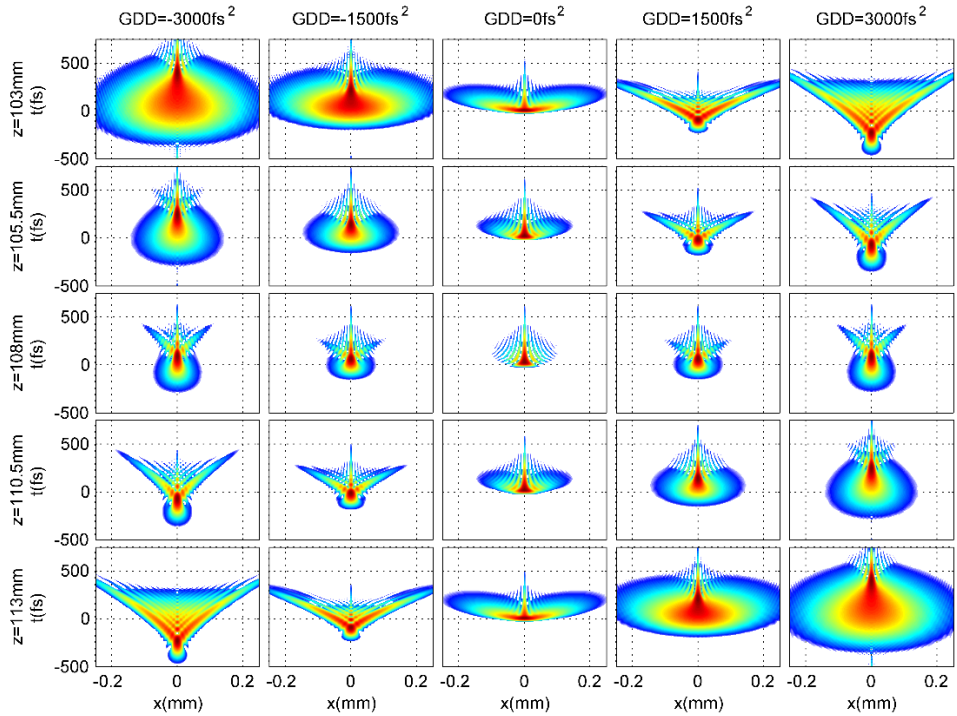

Fig. 3. Theoretical spatio-temporal irradiance distribution depending on the propagation plane z (rows) and GDD of the input pulse (columns). δz=0 mm corresponds to $\lambda_0$ focal plane. Results are normalized and shown in logarithmic scale, representing 3 orders of magnitude.

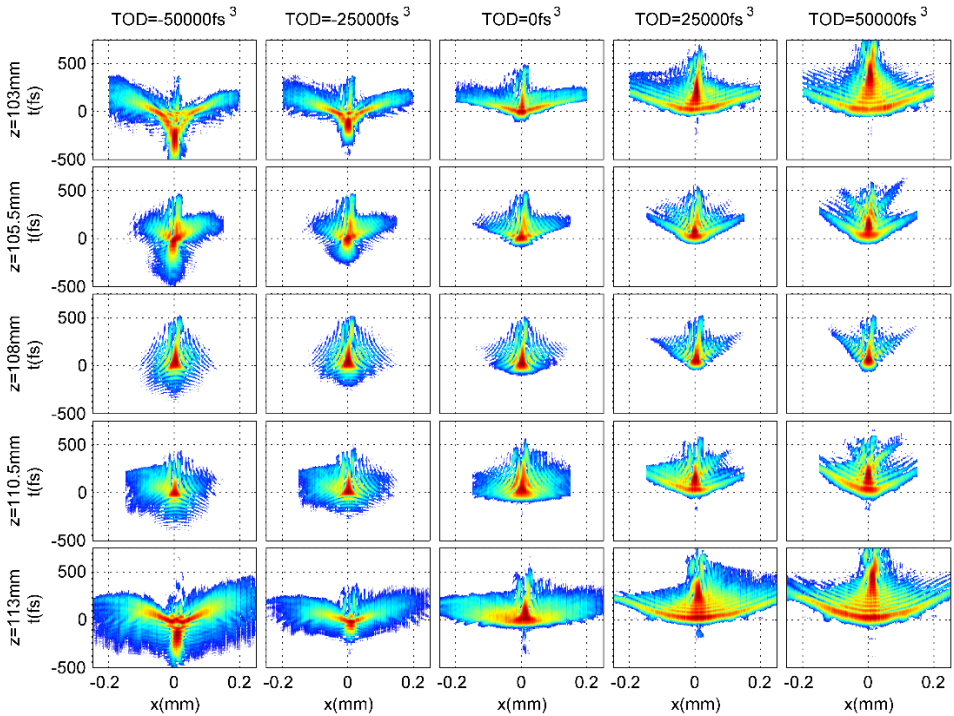

Fig. 4. Experimental spatio-temporal irradiance distribution depending on the propagation plane z (rows) and TOD of the input pulse (columns). δz=0 mm corresponds to $\lambda_0$ focal plane. Results are normalized and shown in logarithmic scale, representing 3 orders of magnitude.

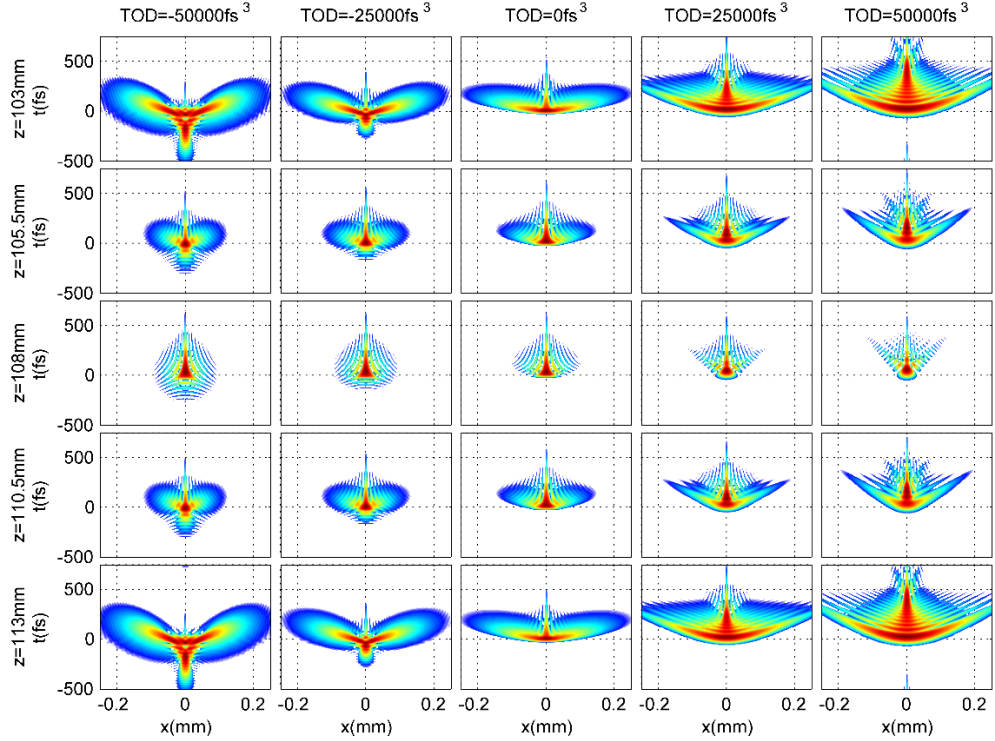

Fig. 5. Theoretical spatio-temporal irradiance distribution depending on the propagation plane z (rows) and TOD of the input pulse (columns). δz=0 mm corresponds to $\lambda_0$ focal plane. Results are normalized and shown in logarithmic scale, representing 3 orders of magnitude.

In a similar way, the cases corresponding to the scan of input pulse TOD (while $GDD = 0\,fs^2$) are shown in Fig. 4 (experiment) and Fig. 5 (simulation). Firstly, the experimental measurements and the simulations, which are in good agreement, show how pulse shaping can alter the spatio-temporal irradiance pattern in the focal region. Secondly, Fig. 4 and Fig. 5 show that also in the TOD scan case there exist different parameter combinations presenting the same spatio-temporal distribution. In this case, contrary to the observed in the GDD case, the $\delta z = 0\,mm$ row acts as symmetry axis: for example, when comparing $\delta z = -5\,mm$ with $\delta z = +5\,mm$, at any value of TOD, e.g. $-50000\,fs^3$, $0\,fs^3$ or $+50000\,fs^3$.

To gain a further intuitive insight on the above commented symmetries, the instantaneous wavelength (i.e., the wavelength associated to the instantaneous frequency calculated from the temporal phase) distributions corresponding to some representative cases from the GDD (Fig. 3) and the TOD scan (Fig. 5) are shown in Fig. 6. The pairs of patterns exhibiting the same spatio-temporal irradiance (e.g., for the GDD scan, $\delta z = +5\,mm$ and $GDD = -3000\,fs^2$ and $\delta z = -5\,mm$ and $GDD = +3000\,fs^2$; for the TOD scan, $\delta z = +5\,mm$ and $TOD = -50000\,fs^3$, and $\delta z = -5\,mm$ and $TOD = -50000\,fs^3$) show inverted instantaneous wavelength with respect to $\lambda_0$. In other words, the regions of the spatio-temporal irradiance distribution exhibiting an instantaneous frequency $\omega_0 + \Omega$ (where the frequency $\omega_0$ is the corresponding to the wavelength $\lambda_0$), appear in the other case of the pair with an instantaneous frequency

$\omega_0 - \Omega$. Therefore, the equivalent cases for both GDD and TOD scans, present equal spatio-temporal irradiance distributions, while the spectral distribution within the pulses is the opposite. This was expected due to the spectral inversion shown in Fig. 1.

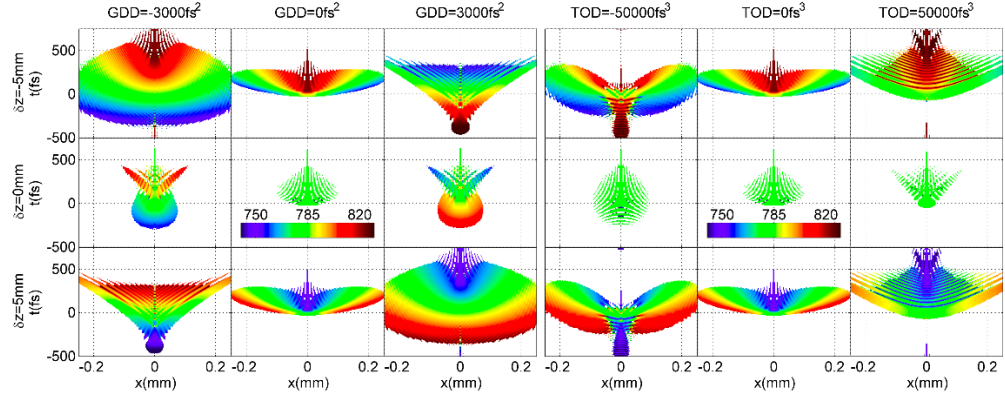

Fig. 6. Theoretical spatio-temporal instantaneous wavelength distribution depending on the propagation plane at δz with respect to f₀ (rows) and the GDD or TOD (columns) phase components added to the input pulse. The color scale bar corresponds to the instantaneous wavelength value in nanometers.

The impact of the propagation position $\delta z$ and the spectral phase component (GDD on TOD) over the instantaneous wavelength distribution can be understood intuitively at some particular cases. One example, observing Fig. 6, is the $\delta z = 0\,mm$ row for the GDD scan case (i.e., the three first columns). In the case of $\delta z = 0\,mm$, $GDD = 0\,fs^2$, the instantaneous wavelength is constant (and equal to the central wavelength) within the whole spatio-temporal distribution. Therefore, because of the symmetrical nature of the second order component of the dispersion, combined with the symmetrical spectrum shown in Fig. 1 ($\delta z = 0$), we can deduce that the addition of a certain amount of GDD (e.g., $GDD = +3000\,fs^2$) will yield the same spatio-temporal irradiance pattern than using the opposite GDD value (e.g., $GDD = -3000\,fs^2$), but with opposite chirp (as seen in $\delta z = 0$ row of Fig. 6).

Concerning the TOD scan (i.e., the three columns on the right of Fig. 6), all the spatio-temporal light distributions at $\delta z = 0\,mm$ show constant instantaneous wavelength (being also here the central wavelength). On the other hand, please note that, as shown in the $GDD = 0\,fs^2$ column, the spatio-temporal light distribution at $+\delta z$ is similar to the found at $-\delta z$ but, again, with opposite chirp (e.g., $\delta z = -5\,mm$ and $\delta z = +5\,mm$ for $GDD = 0\,fs^2$). Therefore, these two elements yield equal spatio-temporal irradiance distributions exhibiting opposite chirp when planes at $+\delta z$ and $-\delta z$ are considered (e.g., $\delta z = \pm 5\,mm$ cases at the diverse TOD columns).

For a more comprehensive insight into the light patterns, in the following section we find an analytical approximation of the Fresnel diffraction expression and we demonstrate the symmetries observed in the spatio-temporal light distributions.

## 4. Discussion

In order to understand the spatio-temporal structures in the region near the focus, we consider the expression of the diffraction given by the Fresnel integral in cylindrical coordinates [18]:

$$U_2(z,\omega;r_2) = \frac{i\omega}{cz}\exp\left(\frac{-i\omega r_2^2}{2cz}\right)\int_0^a U_1(\omega;r_1)\exp\left(\frac{-i\omega r_1^2}{2cz}\right)\exp\left(\frac{i\omega r_1^2}{2cf(\omega)}\right)J_0\left(\frac{\omega r_1 r_2}{cz}\right)r_1 dr_1 \quad (1)$$

where the effect of the DL has been introduced through the term $\exp\left(\frac{i\omega r_1^2}{2cf(\omega)}\right)$ [18].

We express $z = f_0 + \delta z$ and $\omega = \omega_0 + \Omega$, where our assumptions are $\delta z/f_0 \ll 1$ and $\Omega/\omega_0 \ll 1$, which are fulfilled when studying the focusing region. By approximating $\frac{1}{z} = \frac{1}{f_0 + \delta z} \approx \frac{1}{f_0}\left(1 - \frac{\delta z}{f_0}\right)$ and neglecting second order terms, Eq. (1) can be rewritten as

$$U_2(z,\omega;r_2) \approx \frac{i\omega_0}{cf_0}\left(1 - \frac{\delta z}{f_0} + \frac{\Omega}{\omega_0}\right)\exp\left(\frac{-ir_2^2}{2c}\frac{\omega_0}{f_0}\left(1 - \frac{\delta z}{f_0} + \frac{\Omega}{\omega_0}\right)\right) \times$$
$$\int_0^a U_1(\omega;r_1)\exp\left(\frac{ir_1^2}{2c}\frac{\omega_0}{f_0}\left(\frac{\delta z}{f_0} - \frac{\Omega}{\omega_0}\right)\right)J_0\left(\frac{r_1 r_2}{c}\frac{\omega_0}{f_0}\left(1 - \frac{\delta z}{f_0} + \frac{\Omega}{\omega_0}\right)\right)r_1 dr_1 \quad (2)$$

Concerning the factors out of the integral, we can use the zero-order approximation $\frac{\omega_0}{f_0}\left(1 - \frac{\delta z}{f_0} + \frac{\Omega}{\omega_0}\right) \approx \frac{\omega_0}{f_0}$ in the amplitude term. In the case of the term affecting the phase, we do the approximation $\exp\left(\frac{-ir_2^2}{2c}\frac{\omega_0}{f_0}\left(1 - \frac{\delta z}{f_0} + \frac{\Omega}{\omega_0}\right)\right) \approx 1$, taking into account that, apart from the observation conditions close to the focal point ($\delta z/f_0 \ll 1$, $\Omega/\omega_0 \ll 1$), in the case that we are studying, $\frac{r_2^2}{2c}\frac{\omega_0}{f_0} \ll 1$. Finally, as it will be shown in Appendix A, again for those conditions, we can approximate $J_0\left(\frac{r_1 r_2}{c}\frac{\omega_0}{f_0}\left(1 - \frac{\delta z}{f_0} + \frac{\Omega}{\omega_0}\right)\right) \approx J_0\left(\frac{r_1 r_2}{c}\frac{\omega_0}{f_0}\right)$. Then, the Eq. (1) takes the following form, giving the approximated Fresnel diffraction

$$U_2(z,\omega;r_2) \approx \frac{i}{c}\frac{\omega_0}{f_0}\int_0^a U_1(\omega;r_1)\exp\left(\frac{ir_1^2}{2c}\frac{\omega_0}{f_0}\left(\frac{\delta z}{f_0} - \frac{\Omega}{\omega_0}\right)\right)J_0\left(\frac{r_1 r_2}{c}\frac{\omega_0}{f_0}\right)r_1 dr_1 \quad (3)$$

from which we can see that the inversion of the relative propagation distance $\delta z$ and the relative frequency $\Omega = \omega - \omega_0$ is translated into a conjugation of the field (the amplitude is kept, whereas the phase is inverted), provided that the amplitude of the input spectrum $U_1$ is symmetrical with respect to $\omega_0$, this is, $U_1(+\Omega;r_1) = U_1(-\Omega;r_1)$. Notice that for the moment we consider the spectral phase of the input pulse to be zero. Therefore:

$$U_2(-\delta z, -\Omega; r_2) = -U_2^*(\delta z, \Omega; r_2) \quad (4)$$

Please note that the imaginary factor $i$ in the second term of Eq. (3) is responsible of the negative sign under the conjugation in the second term of Eq. (4), consistent with the Gouy phase shift appearing when propagating through the focus.

To calculate the spatio-temporal distribution of the electric field, we apply Fourier transform, denoted by $\mathbb{F}$, and from Eq. (4) we have

$$\mathbb{F}\{U_2(-\delta z,-\Omega;r_2)\} = \int_{-\infty}^{+\infty} U_2(-\delta z,-\Omega;r_2) e^{i\Omega t} d\Omega =$$
$$= -\int_{-\infty}^{+\infty} U_2^*(+\delta z,+\Omega;r_2) e^{i\Omega t} d\Omega = -\left[\mathbb{F}\{U_2(\delta z,\Omega;r_2)\}\right]^* \quad (5)$$

Since there is a conjugation of the electric field expression in the temporal domain (the negative sign denotes a constant $\pi$ phase, affecting only the pulse carrier envelope phase but not the pulse shape), the intensity (amplitude) profile of the pulse is conserved when inverting planes and frequencies, i.e., the roles of the instantaneous frequencies (chirp) are inverted with respect to $\omega_0$, This result explains the behavior observed at the $GDD = 0\,fs^2/TOD = 0\,fs^3$ columns in Fig. 2 to 6.

In the present work, the pulse prior the focusing element is shaped by altering the input spectral phase $\varphi(\omega)$. We can express this phase as a Taylor series $\varphi(\Omega) = \sum_{j=1}^{m} \frac{\varphi^{(j)}}{j!}\Omega^j$. If odd orders of the spectral phase are applied (e.g., the TOD), then we have:

$$\mathbb{F}\left\{e^{i\frac{\varphi^{(2n-1)}}{(2n-1)!}(-\Omega)^{(2n-1)}} U_2(-\delta z,-\Omega;r_2)\right\} = -\left(\mathbb{F}\left\{e^{i\frac{\varphi^{(2n-1)}}{(2n-1)!}\Omega^{(2n-1)}} U_2(\delta z,\Omega;r_2)\right\}\right)^* \quad (6)$$

where n=1,2,3…

On the other hand, for even orders of the spectral phase (e.g., the GDD), we have:

$$\mathbb{F}\left\{e^{i\frac{\varphi^{(2n)}}{(2n)!}(-\Omega)^{2n}} U_2(-\delta z,-\Omega;r_2)\right\} = -\left(\mathbb{F}\left\{e^{i\frac{-\varphi^{(2n)}}{(2n)!}\Omega^{2n}} U_2(\delta z,\Omega;r_2)\right\}\right)^* \quad (7)$$

where n=1,2,3…

Therefore, according to Eq. (6), when only odd order components $\varphi^{(2n-1)}$ of the spectral phase are applied, the amplitude (and irradiance) spatio-temporal pattern is maintained when $\delta z \to -\delta z$ and $\Omega \to -\Omega$. On the other hand, from Eq. (7), in order to preserve those symmetries when only even order components $\varphi^{(2n)}$ of the spectral phase are applied, the sign of such components must be changed. Therefore, this explains the behaviour observed numerically and experimentally in the previous section.

To numerically test the validity of our approximation, we compared the calculation of the spatially-resolved spectrum and the spatio-temporal intensity using the full and the approximated expression for Fresnel diffraction (Eqs. (1) and (3), respectively), at different propagation distances and with similar conditions than in the experiments, being the difference between the results always less than 1.5%.

## 5. Conclusions

In the present work, we have combined diffractive element focusing and temporal shaping (by adding input spectral phase) to alter the pulse structure at the beam focus and the surrounding planes. We have found that it is possible to obtain diverse spatio-temporal distributions of the irradiance (e.g., conical, round, winged, etc.), as well as to control the frequency ordering (temporal chirp) within the pulses. Therefore, pulse shaping is presented as a control parameter to be added to the spatial shaping obtained from DOEs, in order to tailor beams on target and their evolution along the focusing region. Experimental measurements are done with the STARFISH technique, while simulations are done under the scalar diffraction theory, being the results in agreement, thus corroborating our hypothesis. In the case of tight focusing, other theoretical models must be used [29].

We study separately the effect of the GDD and TOD components in analytical, numerical and experimental manner, from which it has been possible to establish the role of those components, and the presence of symmetries on the beam and pulse distribution referred to the considered parameters, i.e., propagation distance and input chirp.

Once the dependences of the spatio-temporal beam distributions have been understood, they present potential for further utilization. For instance, if wavelength dependence of the focal length is inverted, the results are similar but presenting an inversion on the temporal coordinate. Therefore, through the chromaticity it is possible to change the concavity/convexity of the spatio-temporal structure. The combination of both parameters can yield different spatio-temporal structures at a given propagation plane such as X-shape or hollow figures (ellipsoids, double cone, etc.).

## Appendix A

Let's consider $J_0(\alpha(1+h))$, with $h \ll 1$ and $\alpha h \ll 1$, as it happens in the experimental conditions of the present work. Since

$$J_0(\alpha) = \frac{1}{\pi}\int_0^\pi \cos(\alpha \sin(\theta))d\theta \tag{8}$$

then,

$$\begin{aligned} J_0(\alpha(1+h)) &= \frac{1}{\pi}\int_0^\pi \cos(\alpha \sin(\theta)(1+h))d\theta = \\ &= \frac{1}{\pi}\int_0^\pi \left[\cos(\alpha \sin(\theta))\cos(h\alpha \sin(\theta)) - \sin(\alpha \sin(\theta))\sin(h\alpha \sin(\theta))\right]d\theta \approx \\ &\approx \frac{1}{\pi}\left[\int_0^\pi \cos(\alpha \sin(\theta))d\theta - h\alpha\int_0^\pi \sin(\theta)\sin(\alpha \sin(\theta))d\theta\right] = \\ &= J_0(\alpha) - \frac{1}{\pi}h\alpha\int_0^\pi \sin(\theta)\sin(\alpha \sin(\theta))d\theta \approx J_0(\alpha) \end{aligned} \tag{9}$$

Considering the parameters linked to the presented experiments, the error of this approximation is around 1-2% maximum.

## Funding

Junta de Castilla y León (SA046U16); Spanish Ministerio de Economía y Competitividad (MINECO) (FIS2015-71933-REDT, FIS2016-75618-R, FIS2017-87970-R); Generalitat Valenciana (PROMETEU/2016/079); University Jaume I (UJI-B2016-19); European Union (798264).

## Acknowledgments

The Centro de Láseres Pulsados (CLPU) is acknowledged for granting access to its facility. E. Conejero-Jarque is acknowledged for fruitful discussions. B.A. acknowledges funding from the European Union's Horizon 2020 research and innovation programme under the Marie Sklodowska-Curie Individual Fellowship grant agreement No. 798264, project FastMeasure entitled 'Development and industry transfer of new techniques: full characterization of vector ultrashort pulsed laser beams'.